\documentclass[sigconf]{acmart}
\usepackage[T1]{fontenc}
\usepackage{lmodern}
\usepackage{algorithm}
\usepackage{algorithmicx}
\usepackage{algpseudocode}
\usepackage{amsmath}
\usepackage{amssymb}
\usepackage{array}
\usepackage{arydshln}
\usepackage{bm}
\usepackage{booktabs}
\usepackage{braket}
\usepackage{enumitem}
\usepackage{graphicx}
\usepackage[mathscr]{euscript}
\usepackage{mathtools}
\usepackage{multirow}
\usepackage{framed} 
\usepackage{nicefrac}
\usepackage{pgfplots}
\usepackage{qtree}
\usepackage{adjustbox}
\usepackage{rotating}
\pgfplotsset{compat = 1.13}
\usepackage{stfloats}
\usepackage{subfig}
\usepackage{url}
\usepackage{tikz}
\usepackage{tkz-berge}
\usetikzlibrary{shapes, backgrounds,calc}
\usetikzlibrary{decorations.pathreplacing}
\usepackage{fancybox}

\tikzstyle{vertex} = [circle, draw, inner sep = 0pt, minimum size = 10pt]
\newcommand{\vertex}{\node[vertex]}

\definecolor{bblue}{HTML}{2900DD}
\definecolor{rred}{HTML}{EA00F4}
\definecolor{ggreen}{HTML}{33F413}
\definecolor{yyellow}{HTML}{F4BF05}

\setcopyright{rightsretained}

\copyrightyear{2017} 
\acmYear{2017} 
\setcopyright{acmcopyright}
\acmConference{MobiCom'17}{}{October 16--20, 2017, Snowbird, UT, USA.}
\acmPrice{15.00}
\acmDOI{http://dx.doi.org/10.1145/3117811.3131260}
\acmISBN{ISBN 978-1-4503-4916-1/17/10} 

\copyrightyear{2017} 
\acmYear{2017} 
\setcopyright{acmlicensed}
\acmConference{MobiCom'17}{}{October 16--20, 2017, Snowbird, UT, USA.}
\acmPrice{15.00}
\acmDOI{http://dx.doi.org/10.1145/3117811.3131260}
\acmISBN{ISBN 978-1-4503-4916-1/17/10} 

\copyrightyear{2017} 
\acmYear{2017} 
\setcopyright{rightsretained}
\acmConference{MobiCom'17}{}{October 16--20, 2017, Snowbird, UT, USA.}
\acmPrice{} 
\acmDOI{http://dx.doi.org/10.1145/3117811.3131260}
\acmISBN{ISBN 978-1-4503-4916-1/17/10}

\begin{document}
\fancyhead{}
\settopmatter{printacmref=false, printfolios=false}

\title{Poster: Resource Allocation with Conflict Resolution for Vehicular Sidelink Broadcast Communications}

\author{Luis F. Abanto-Leon}
\affiliation{%
  \institution{Eindhoven University of Technology}
  \city{Eindhoven} 
  \country{Netherlands} 
}
\email{l.f.abanto@tue.nl}

\author{Arie Koppelaar}
\affiliation{%
	\institution{NXP Semiconductors}
	\city{Eindhoven} 
	\country{Netherlands} 
}
\email{arie.koppelaar@nxp.com}

\author{Sonia Heemstra de Groot}
\affiliation{%
	\institution{Eindhoven University of Technology}
	\city{Eindhoven} 
	\country{Netherlands} 
}
\email{sheemstradegroot@tue.nl}

\renewcommand{\shortauthors}{L. F. Abanto-Leon et al.}

\begin{abstract}
	In this paper we present a graph-based resource allocation scheme for sidelink broadcast V2V communications. Harnessing available information on geographical position of vehicles and spectrum resources utilization, eNodeBs are capable of allotting the same set of sidelink resources to different vehicles distributed among several communications clusters. Within a communications cluster, it is crucial to prevent time-domain allocation conflicts since vehicles cannot transmit and receive simultaneously, i.e., they must transmit in orthogonal time resources. In this research, we present a solution based on a bipartite graph, where vehicles and spectrum resources are represented by vertices whereas the edges represent the achievable rate in each resource based on the SINR that each vehicle perceives. The aforementioned time orthogonality constraint can be approached by aggregating conflicting vertices into macro-vertices which, in addition, reduces the search complexity. We show mathematically and through simulations that the proposed approach yields an optimal solution. In addition, we provide simulations showing that the proposed method outperforms other competing approaches, specially in scenarios with high vehicular density. 
\end{abstract}

%
%


\copyrightyear{2017} 
\acmYear{2017} 
\setcopyright{rightsretained} 
\acmConference{MobiCom '17}{October 16--20, 2017}{Snowbird, UT, USA}\acmDOI{10.1145/3117811.3131260}
\acmISBN{978-1-4503-4916-1/17/10}

\maketitle
\section{Introduction}
Vehicle--to--vehicle (V2V) communications is one of the novel use cases in 5G and has attracted much interest, specially for safety applications, since connected vehicles may have the potential to prevent accidents \cite{b1}.

In V2V $\textit{Mode 3}$, eNodeBs assign resources to vehicles for them to periodically broadcast CAM messages \cite{b2}. Once the allocation has been accomplished, data is disseminated directly between vehicles. Conversely, in conventional cellular communications, data traverse the eNodeB via uplink/downlink before it can be forwarded. Thus, since data traffic is controlled by the eNodeB, users can be allocated in the same time resource but different frequency subchannels. Nevertheless, in V2V $\textit{Mode 3}$ the flow of data is not controlled, thus it is fundamental to guarantee that vehicles will transmit in orthogonal time resources to prevent conflicts. On the other hand, assignment problems can be represented as weighted bipartite graphs, where the objective is to find a maximal matching. This classical problem is called herein \textit{unconstrained weighted graph matching}. Resource allocation for V2V communications has a time orthogonality constraint which cannot be handled by the mentioned method. We have therefore envisaged a solution called \textit{constrained weighted graph matching}, which incorporates the mentioned constraint.

The objective of this paper is two-fold: $\big( i \big)$ prove that an optimal solution for the \textit{constrained weighted graph matching} problem exists and $\big( ii \big)$ discuss the suitability of such an approach for avoiding resource allocation conflicts in broadcast vehicular communications.

\section{Motivation and Contributions}
Our motivation is to develop an approach capable of $\big( a \big)$ preventing allocation conflicts---by enforcing constraints---and $\big( b \big)$ maximizing the sum-rate capacity of the system. For instance, in one of the clusters of Fig. 1, a resource conflict can be observed between vehicles $V_8$ and $V_{10}$ since they have been allotted resources in the same time subframe. 
\begin{figure}[t]
	\begin{center}
		\includegraphics[width=0.90\linewidth]{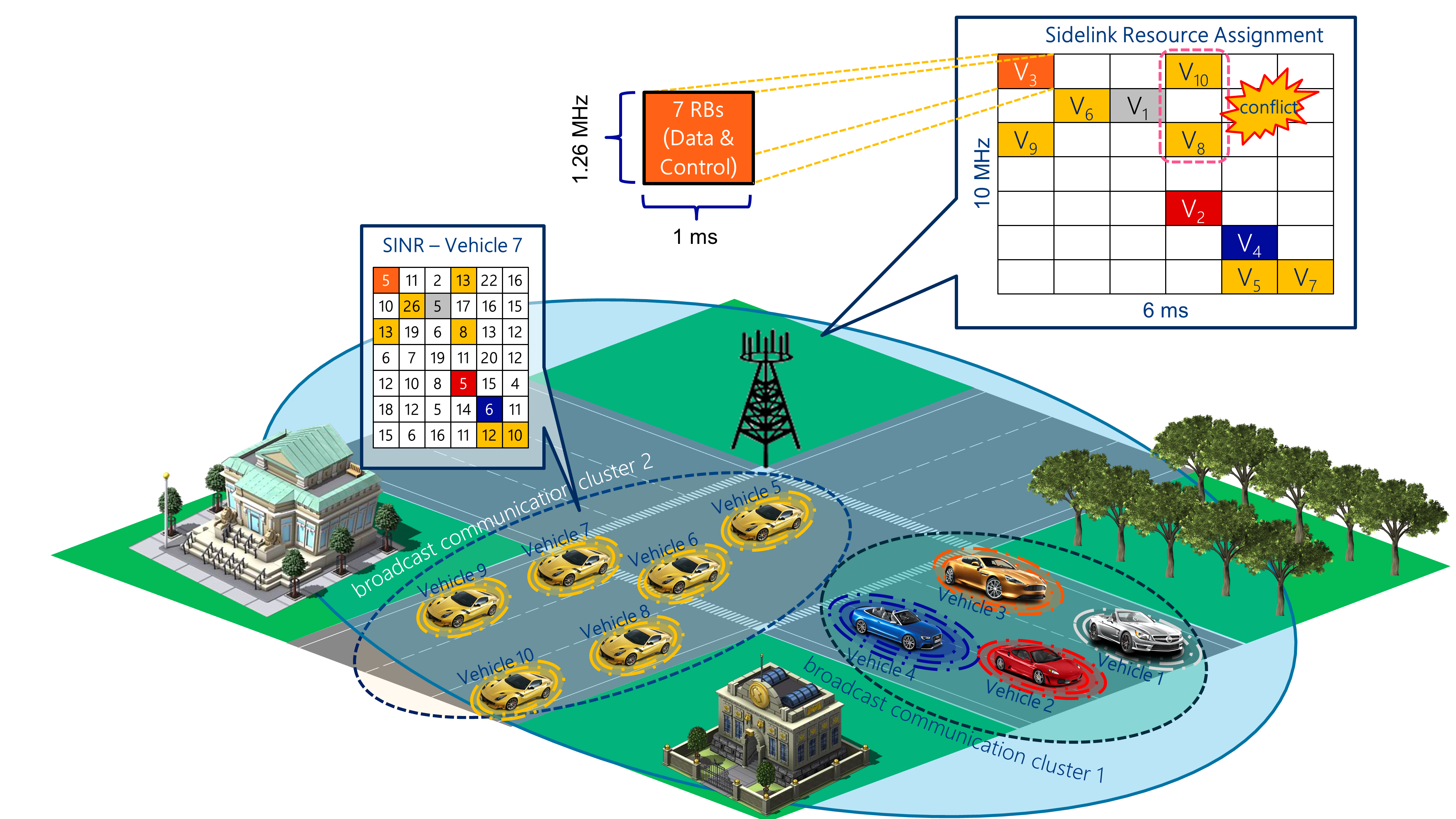}
	\end{center}
	\caption{V2V Broadcast Communications via Sidelink}
	\label{Fig1}
	\vspace{-0.2cm}
\end{figure}

The contributions of our work are summarized:
\begin{itemize}
	\item Kuhn-Munkres \cite{b3} is a computationally efficient method for solving matching problems in bipartite graphs. However, due to additional time orthogonality constraints, the resultant problem is not directly approachable by it. In our solution, vertices conflicting among each other have been aggregated into macro-vertices yielding a resultant graph which is solvable by Kuhn-Munkres.
	\item Vertex aggregation cuts down the number of effective vertices and therefore narrows the amount of potential solutions, without affecting optimality.
	\item We show through simulations that our approach is capable of providing fairness among all vehicles, especially in scenarios with high vehicle density.
\end{itemize}

\section{Proposed Approach}
Let $G = \big( \mathcal{V}, \mathcal{R}, \mathcal{E} \big)$ be a bipartite graph such that $\vert\mathcal{R}\vert = K \vert\mathcal{V}\vert = KN$, as depicted in Fig. 2. In this scheme, the $KN$ vertices in $\mathcal{R}$ are clustered into $N$ disjoint groups $\{\mathcal{R}_{\alpha}\}_{\alpha = 1}^N$ called macro-vertices, such that $\mathcal{R} = \cup_{\alpha = 1}^{N} \mathcal{R}_{\alpha}$, $\mathcal{R}_{\alpha} \cap \mathcal{R}_{\alpha'} = \emptyset$, $\forall \alpha \neq \alpha'$. Each macro-vertex $\mathcal{R}_{\alpha}$ is an aggregation of $K$ vertices, i.e., $\vert\mathcal{R}_{\alpha}\vert = K$. The target is to find a vertex--to--vertex (or vehicle--to--resource) matching with maximum sum of weights such that no two vertices in $\mathcal{V}$ are matched to any two vertices in the same macro-vertex $\mathcal{R}_{\alpha}$. This condition must be satisfied as it depicts the time orthogonality requirement that prevents allocation conflicts. We will show that the optimal solution is tantamount to finding the maximum vertex--to--macro-vertex matching. 

In Fig. 2, vertices $v_i$ represent the vehicles in cluster $\mathcal{V}$ whereas vertices $r_j$ represent the allotable resources managed by the eNodeB. $K$ represents the number of resources per subframe. $N$ represents the amount of available subframes in which the resource allocation task can be accomplished. The problem is formulated by (\ref{e1}) 
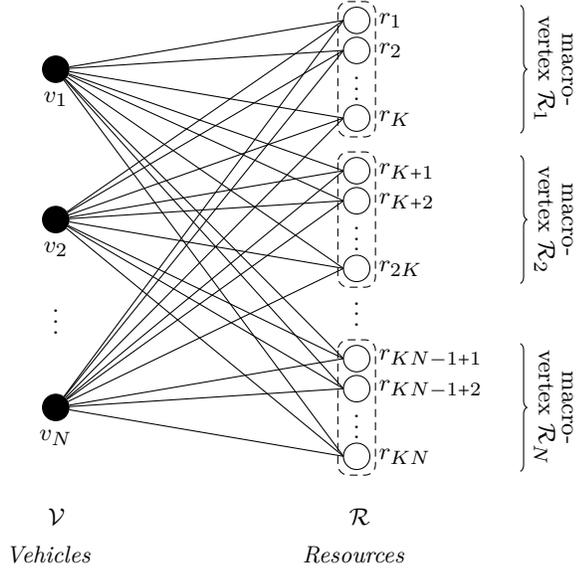
\begin{figure}[!t]
	\centering
	\[\begin{tikzpicture}
	
	\vertex[fill] (v1) at (0,-1.25) [label = below:$v_{1}$] {};
	\vertex[fill] (v2) at (0,-3.25) [label = below:$v_{2}$] {};
	\node at (0,-4.5) {\vdots};
	\vertex[fill] (v3) at (0,-5.75) [label = below:$v_{N}$] {};
	
	\vertex (r11) at (4,-0.6) [label = right:$r_{1}$] {};
	\vertex (r12) at (4,-1) [label = right:$r_{2}$] {};
	\node at (4,-1.4) {\vdots};
	\vertex (r13) at (4,-1.9) [label = right:$r_{K}$] {};
	
	\vertex (r21) at (4,-2.6) [label = right:$r_{K+1}$] {};
	\vertex (r22) at (4,-3) [label = right:$r_{K+2}$] {};
	\node at (4,-3.4) {\vdots};
	\vertex (r23) at (4,-3.9) [label = right:$r_{2K}$] {};
	
	\node at (4,-4.4) {\vdots};
	
	\vertex (r31) at (4,-5.1) [label = right:$r_{K(N-1)+1}$] {};
	\vertex (r32) at (4,-5.5) [label =right:$r_{K(N-1)+2}$] {};
	\node at (4,-5.9) {\vdots};
	\vertex (r33) at (4,-6.4) [label = right:$r_{KN}$] {};
	
	\path
	(v1) edge (3.825,-0.6)
	(v1) edge (3.825,-1)
	(v1) edge (3.825,-1.9)
	(v1) edge (3.825,-2.6)
	(v1) edge (3.825,-3)
	(v1) edge (3.825,-3.9)
	(v1) edge (3.825,-5.1)
	(v1) edge (3.825,-5.5)
	(v1) edge (3.825,-6.4)
	
	(v2) edge (3.825,-0.6)
	(v2) edge (3.825,-1)
	(v2) edge (3.825,-1.9)
	(v2) edge (3.825,-2.6)
	(v2) edge (3.825,-3)
	(v2) edge (3.825,-3.9)
	(v2) edge (3.825,-5.1)
	(v2) edge (3.825,-5.5)
	(v2) edge (3.825,-6.4)
	
	(v3) edge (3.825,-0.6)
	(v3) edge (3.825,-1)
	(v3) edge (3.825,-1.9)
	(v3) edge (3.825,-2.6)
	(v3) edge (3.825,-3)
	(v3) edge (3.825,-3.9)
	(v3) edge (3.825,-5.1)
	(v3) edge (3.825,-5.5)
	(v3) edge (3.825,-6.4);
	
	\draw[densely dashed,rounded corners=4]($(r11)+(-.25,.25)$)rectangle($(r13)+(0.25,-.25)$);
	\draw[densely dashed,rounded corners=4]($(r21)+(-.25,.25)$)rectangle($(r23)+(0.25,-.25)$);
	\draw[densely dashed,rounded corners=4]($(r31)+(-.25,.25)$)rectangle($(r33)+(0.25,-.25)$);
	
	\draw[decoration={brace, raise=5pt},decorate] (6,-0.4) -- node[right=6pt] {} (6,-2.1);
	\draw[decoration={brace, raise=5pt},decorate] (6,-2.4) -- node[right=6pt] {} (6,-4.1);
	\draw[decoration={brace, raise=5pt},decorate] (6,-4.9) -- node[right=6pt] {} (6,-6.6);
	
	\node[rotate=-90] at (6.8,-1.25) {macro-};
	\node[rotate=-90] at (6.8,-3.25) {macro-};
	\node[rotate=-90] at (6.8,-5.75) {macro-};
	\node[rotate=-90] at (6.5,-1.25) {vertex $\mathcal{R}_1$};
	\node[rotate=-90] at (6.5,-3.25) {vertex $\mathcal{R}_2$};
	\node[rotate=-90] at (6.5,-5.75) {vertex $\mathcal{R}_N$};
	
	\node[text width = 0.2cm] at (0,-7.2) {$\mathcal{V}$};
	\node[text width = 0.2cm] at (-0.5,-7.7) {\textit{Vehicles}};
	\node[text width = 0.2cm] at (4,-7.2) {$\mathcal{R}$};
	\node[text width = 0.2cm] at (3.4,-7.7) {\textit{Resources}};
	
	\end{tikzpicture}\]
	\caption{Constrained Weighted Bipartite Graph}
	\label{Fig2}
	\vspace{-0.45cm}
\end{figure}
\begin{equation} \label{e1}
\hspace{-0.87cm}
\begin{array}{lclcl}
&& {\rm max} ~ {\bf c}^T {\bf x} \\
&& {\rm subject~to}~ 
	\underbrace{
		{\left[
			\begin{array}{c}
			{\bf I}_{N \times N} \otimes {\bf 1}_{1 \times N}\\
			\hline
			{\bf 1}_{1 \times N} \otimes {\bf I}_{N \times N} 
			\end{array}
			\right]}}_\text{\bf A} \otimes {\bf 1}_{1 \times K}~ {\bf x} = {\bf 1}
\end{array}
\end{equation}
where $\otimes$ represents the tensor product operator, ${\bf c} \in \mathbb{R}^{M},  {\bf x} \in \mathbb{B}^M$ with $M = KN^2$. The solution and weight vectors are $\mathbf{x} = \big[  x_{1,1}, \dots,  x_{N,N} \big] ^T$ and $\mathbf{c} = \big[ c_{1,1}, \dots, c_{N,N} \big] ^T$, respectively. Also, $c_{ij} = B \log_2 \big( 1 + \mathsf{SINR}_{ij} \big)$. Because $\bf x$ exists on the binary subspace, the cost function can be equivalently expressed as ${\bf c}^T {\bf x} = {\bf x}^T diag \big( {\bf c} \big)  {\bf x}$ without affecting optimality. Also, we can add zero-valued terms $c_{ij} x_{ij} x_{ik}$ (with $r_j, r_k \in \mathcal{R}_{\alpha}$) to the cost function without affecting the solution. A more generalized expression is given by  ${\bf x}^T \big( {\bf I}_{M \times M} \otimes [{\bf 1}_{K \times K} - {\bf I}_{K \times K}] \big) diag \big( {\bf c} \big) {\bf x} = 0.$

\begin{figure}[!b]
	\centering
	\begin{tikzpicture}
	
	\draw (1,0) rectangle (3.5, -0.6) node[pos=.5] {${\bf I}_{M \times M}\otimes {\bf 1}_{1 \times K}$};
	\draw (1,-1) rectangle (3.5, -1.5) node[pos=.5] {${\bf I}_{M \times M}\otimes {\bf 1}_{1 \times K}$};
	\draw (-0.8,-1.1) rectangle (-0.5, -1.4) node[pos=.5] {$\times$};
	\draw (-2.5,-1) rectangle (-1.3, -1.5) node[pos=.5] {$diag(\cdot)$};
	
	\draw [->] (-3, -0.3) -- (1, -0.3);
	\draw [->] (-0.5, -1.25) -- (1, -1.25);
	\draw [->] (-0.65, -0.3) -- (-0.65, -1.1);
	\draw [->] (-1.3, -1.25) -- (-0.8, -1.25);
	\draw [->] (-3, -1.25) -- (-2.5, -1.25);
	\draw [->] (3.5, -0.3) -- (4, -0.3);
	\draw [->] (3.5, -1.25) -- (4, -1.25);
	
	\node[text width = 0.2cm] at (-3.2,-0.3) {$\bf x$};
	\node[text width = 0.2cm] at (-3.2,-1.25) {$\bf c$};
	\node[text width = 0.2cm] at (4.2,-0.3) {$\bf y$};
	\node[text width = 0.2cm] at (4.2,-1.25) {$\bf d$};
	
	\end{tikzpicture}
	\caption{Transformation Process}
	\vspace{-0.45cm}
\end{figure}
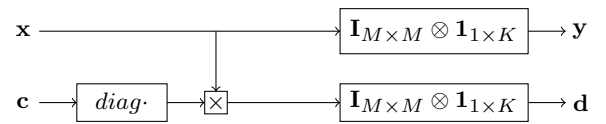
Now, the augmented cost function can be recast as 
\begin{equation} \label{e2}
\small
\hspace{0.1cm}
\begin{array}{l}
~ {\bf c}^T {\bf x} \\

\vspace{-0.2cm} \\

= {\bf x}^T  diag \big( {\bf c} \big) {\bf x} + {\bf x}^T \big(  {\bf I}_{M \times M} \otimes \big[ {\bf 1} - {\bf I} \big]_{K \times K}) \big) ~  diag \big( {\bf c} \big) {\bf x}\\

\vspace{-0.2cm} \\

= {\bf x}^T \big( {\bf I}_{M \times M} \otimes {\bf 1}_{K \times K} \big) diag \big( {\bf c} \big) {\bf x}\\

= {\bf x}^T \big({\bf I}_{M \times M} {\bf I}_{M \times M} \otimes {\bf 1}_{K \times 1} {\bf 1}_{1 \times K}\big) diag \big( {\bf c} \big) {\bf x}\\

\vspace{-0.2cm} \\

= \underbrace{{\bf x}^T \big({\bf I}_{M \times M} \otimes {\bf 1}_{K \times 1}\big)}_{\text{\bf y}^T} 
\underbrace{\big({\bf I}_{M \times M} \otimes {\bf 1}_{1 \times K}\big) diag \big( {\bf c} \big) {\bf x}}_{\text{\bf d}} \\

\end{array}
\end{equation}

From (\ref{e2}), we obtain that ${\bf d} = \big({\bf I}_{M \times M} \otimes {\bf 1}_{1 \times K}\big) diag \big( {\bf c} \big) {\bf x}$ and ${\bf y} = \big({\bf I}_{M \times M} \otimes {\bf 1}_{1 \times K}\big) {\bf x}$. Similarly, we obtain that ${\bf x} = \big({\bf I}_{M \times M} \otimes {\bf 1}_{1 \times K}\big)^{\dagger}{\bf y}$. In the following, we use the previous relations to simplify the constraint in (\ref{e1}),
\begin{equation} \label{e3}
\hspace{0.cm}
\small
\begin{array}{l}

~\left( \left[
\begin{array}{c}
{\bf I}_{N \times N} \otimes {\bf 1}_{1 \times N}\\
\hline
{\bf 1}_{1 \times N} \otimes {\bf I}_{N \times N} 
\end{array}
\right] 
\otimes {\bf 1}_{1 \times K} \right) {\left( {\bf I}_{M \times M} \otimes {\bf 1}_{1 \times K}^{\dagger} \right){\bf y}} = {\bf 1}

\vspace{0.3cm} \\

= \left( \left[
\begin{array}{c}
{\bf I}_{N \times N} \otimes {\bf 1}_{1 \times N}\\
\hline
{\bf 1}_{1 \times N} \otimes {\bf I}_{N \times N} 
\end{array}
\right] 
{\bf I}_{M \times M} \right) \otimes \underbrace{{\left( {\bf 1}_{1 \times K} {\bf 1}_{1 \times K}^{\dagger} \right)}}_\text{1} {\bf y} = {\bf 1}

\vspace{0.01cm} \\

= \left[
\begin{array}{c}
{\bf I}_{N \times N} \otimes {\bf 1}_{1 \times N}\\
\hline
{\bf 1}_{1 \times N} \otimes {\bf I}_{N \times N} 
\end{array}
\right] 
{\bf y} = {\bf 1}

\end{array}
\end{equation}
Thus, the problem in (\ref{e3}) can be recast as (\ref{e4})
\begin{equation} \label{e4}
\hspace{-1.5cm}
\begin{array}{lclcl}
&& {\rm max} ~ {\bf d}^T {\bf y} \\
&& {\rm subject~to}~ 
\underbrace{
	{\left[
		\begin{array}{c}
		{\bf I}_{N \times N} \otimes {\bf 1}_{1 \times N}\\
		\hline
		{\bf 1}_{1 \times N} \otimes {\bf I}_{N \times N} 
		\end{array}
		\right]}}_\text{\bf {A}} {\bf y} = {\bf 1}.
\end{array}
\end{equation}

Fig.3 shows the transformation process from (\ref{e1}) to (\ref{e4}). We notice that $\bf d$ depends on $\bf x$ which is not desirable. In order to eliminate this dependency, we state without a proof---due to space limitations---that
\begin{equation} \label{e18}
{\bf d} = \lim_{\beta \to \infty} \frac{1}{\beta} \overset{\substack{\circ}}{\log} \bigg\{({\bf I}_{M \times M}\otimes {\bf 1}_{1 \times K}) \mathrm{e}^{\circ \beta {\bf c}} \bigg\}
\end{equation}
where $\overset{\substack{\circ}}{\log} \{\cdot\}$ and $\mathrm{e}^{\circ \{ \cdot \} }$ are the element-wise natural logarithm and Hadamard exponential \cite{b4}, respectively.

\section{Simulations}
We consider a 10 MHz channel which is divided into several resource chunks, each with an extent of 1ms in time and 1.26 MHz in frequency. To wit, 1.26 MHz corresponds to 7 resource blocks (RBs), each consisting of data (5RBs) and control information (2RBs) \cite{b5}. In our model, we consider that clusters are independent from each other. Thus, resources used in a certain cluster can be repurposed by vehicles in other clusters. Since we consider a message rate of 10 Hz, the resource allocation task is carried out every 0.1 s and therefore the maximum number of allotable subframes is 100. We also assume that the sidelink channel conditions of each vehicle are reported to the eNodeB via uplink. The resource allocation is broadcasted to vehicles via downlink. Notice that sidelink resources are exclusively used for communications.

In Fig. 4, we compare 4 different algorithms in base of the average over 1000 simulations. We have considered $N = 100$ vehicles in each cluster. Through simulations, we show that our scheme is optimal since it achieves the same performance as exhaustive search. The greedy algorithm performs as good as the proposed approach when we examine the highest-rate vehicle only. This is logical as the premise of the greedy algorithm is to assign the best resources on first-come first-served basis. Considering the system average rate, our proposed approach has an advantage over the greedy algorithm. Also, when considering the worst-rate vehicle, our proposal excels as it is capable of providing a higher level of fairness. In all cases, the random allocation algorithm is outperformed by the other approaches. 
\begin{figure}[h]
	\centering
	\begin{tikzpicture}
	\begin{axis}[
	ybar,
	ymin = 0,
	ymax = 9.5,
	width = 8.5cm,
	height = 6.9cm,
	bar width = 10pt,
	tick align = inside,
	x label style={align=center, font=\footnotesize,},
	ylabel = {Rate [Mbits / s / resource]},
	y label style={at={(-0.075,0.5)}, font=\footnotesize,},
	nodes near coords,
	every node near coord/.append style={color = white, rotate = 90, anchor = east, font = \fontsize{6}{6}\selectfont},
	nodes near coords align = {vertical},
	symbolic x coords = {Highest-Rate Vehicle, Worst-Rate Vehicle, System Average Rate, System Rate Standard Deviation},
	x tick label style = {text width = 1.6cm, align = center, font = \footnotesize,},
	xtick = data,
	enlarge y limits = {value = 0.15, upper},
	enlarge x limits = 0.18,
	legend columns=2,
	legend pos = north east,
	legend style={font=\fontsize{6}{5}\selectfont, text width=2.7cm,text height=0.02cm,text depth=.ex, fill = none, }]
	]
	\addplot[fill = bblue] coordinates {(Highest-Rate Vehicle,  8.97) (Worst-Rate Vehicle, 7.12) (System Average Rate, 8.22) (System Rate Standard Deviation, 1.16)}; \addlegendentry{Exhaustive Search}
	\addplot[fill = rred] coordinates {(Highest-Rate Vehicle, 8.97) (Worst-Rate Vehicle, 7.12) (System Average Rate, 8.22) (System Rate Standard Deviation, 1.16)}; \addlegendentry{Graph-based Algorithm}
	\addplot[fill = ggreen] coordinates {(Highest-Rate Vehicle, 8.91) (Worst-Rate Vehicle, 5.85) (System Average Rate, 8.02) (System Rate Standard Deviation, 1.25)}; \addlegendentry{Greedy Algoritm}
	\addplot[fill = yyellow] coordinates {(Highest-Rate Vehicle, 7.63) (Worst-Rate Vehicle, 1.76) (System Average Rate, 4.52) (System Rate Standard Deviation, 1.67)}; \addlegendentry{Random Algorithm}
	
	\end{axis}
	\end{tikzpicture}
	\caption{Vehicles Data Rate}
	\vspace{-0.1cm}
\end{figure}
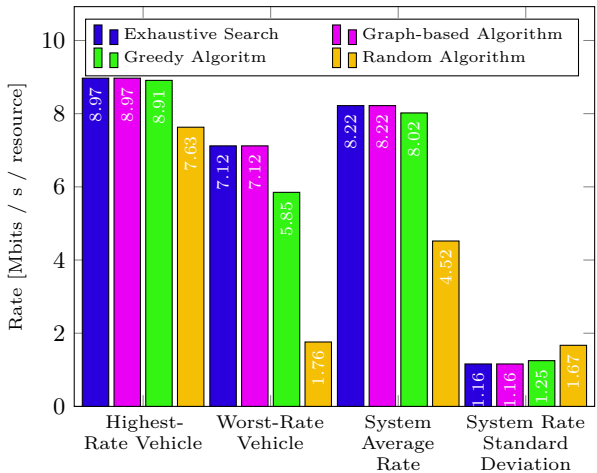
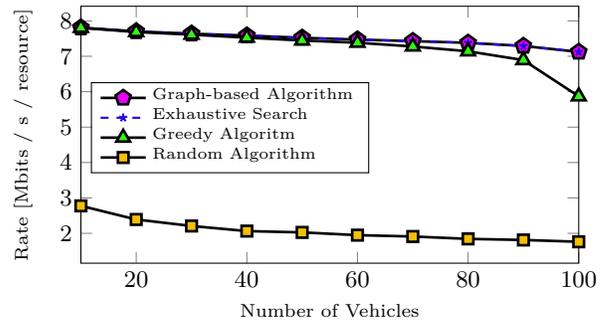
\begin{figure}[!]
	\centering
	\begin{tikzpicture}
	\begin{axis}[
	xmin = 10,
	xmax = 100,
	width = 8.2cm,
	height = 5cm,
	xlabel={Number of Vehicles},
	x label style={align=center, font=\footnotesize,},
	ylabel = {Rate [Mbits / s / resource]},
	y label style={at={(-0.08,0.5)}, text width = 4cm, align=center, font=\footnotesize,},
	ytick = {1, 2, 3, 4, 5, 6, 7, 8},
	legend style={at={(0.02,0.35)},anchor=south west, font=\fontsize{6}{5}\selectfont, text width=2.7cm,text height=0.075cm,text depth=.ex, fill = none,},
	]
	\addplot[color=black, mark = pentagon*, mark options = {scale = 1.5, fill = rred}, line width = 1pt] coordinates 
	{
		(10, 7.8084)
		(20, 7.7028)
		(30, 7.6374)
		(40, 7.5846)
		(50, 7.5216)
		(60, 7.4712)
		(70, 7.4280)
		(80, 7.3794)
		(90, 7.2912)
		(100, 7.1238)
	}; \addlegendentry{Graph-based Algorithm}
	
	\addplot[color=bblue, mark = star, mark options = {scale = 0.8}, line width = 0.8pt, style = dashed] coordinates 
	{
		(10, 7.8084)
		(20, 7.7028)
		(30, 7.6374)
		(40, 7.5846)
		(50, 7.5216)
		(60, 7.4712)
		(70, 7.4280)
		(80, 7.3794)
		(90, 7.2912)
		(100, 7.1238)
	}; \addlegendentry{Exhaustive Search}
	
	\addplot[color=black, mark = triangle*, mark options = {scale = 1.5, fill = ggreen}, line width = 1pt] coordinates 
	{
		(10, 7.8030)
		(20, 7.6848)
		(30, 7.6104)
		(40, 7.5264)
		(50, 7.44545)
		(60, 7.3830)
		(70, 7.2768)
		(80, 7.1400)
		(90, 6.8922)
		(100, 5.8686)
	}; \addlegendentry{Greedy Algoritm}
	
	\addplot[color=black, mark = square*, mark options = {fill = yyellow}, line width = 1pt] coordinates 
	{
		(10, 2.7762)
		(20, 2.3910)
		(30, 2.2074)
		(40, 2.0634)
		(50, 2.0262)
		(60, 1.9488)
		(70, 1.9104)
		(80, 1.8432)
		(90, 1.8126)
		(100, 1.7616)
	}; \addlegendentry{Random Algorithm}
	\end{axis}
	\end{tikzpicture}
	\caption{Worst-rate Vehicle}
	\vspace{-0.1cm}
\end{figure}
Fig. 5 shows the achievable rate for the worst-rate vehicle. The proposed graph-based algorithm attains the same performance as the exhaustive search. We observe that when the vehicle density per cluster is low, the greedy approach attains near optimal solutions as there are far more resources than vehicles to serve. However, as the density increases, especially near the overload state, its performance drops. The random allocation algorithm performs worse than the other approaches.

Fig. 6 shows the CDF of the achievable rates. We observe that the proposed approach outperforms the other two approaches. For the sake of comparison, we have included the results of the unconstrained system, which does not takes into account conflict avoidance constraints. This is of course not desirable but it serves as a comparison bound.
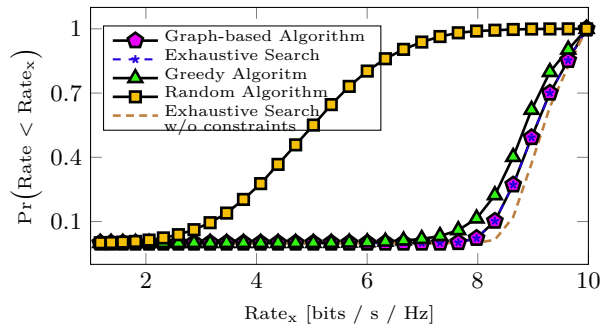
\begin{figure}[!]
	\centering
	\begin{tikzpicture}
	\begin{axis}[
	xmin = 1,
	xmax = 10,
	width = 8.2cm,
	height = 5cm,
	xlabel={Rate\textsubscript{x} [bits / s / Hz]},
	x label style={align=center, font=\footnotesize,},
	ylabel = {Pr\big(Rate~\textless~Rate\textsubscript{x}\big)},
	y label style={at={(-0.08,0.5)}, text width = 3.5cm, align=center, font=\footnotesize,},
	ytick = {0.1, 0.4, 0.7, 1.0},
	legend style={at={(0.025,0.52)},anchor=south west, font=\fontsize{6}{5}\selectfont, text width=2.7cm,text height=0.06cm,text depth=.ex, fill = none, align = left},
	]
	
	\addplot[color = black, mark = pentagon*, mark options = {scale = 1.5, fill = rred, solid}, line width = 1pt] coordinates 
	{
		(1.1756, 0)
		(1.3701, 0)
		(1.5827, 0)
		(1.8122, 0)
		(2.0574, 0)
		(2.3165, 0)
		(2.5878, 0)
		(2.8698, 0)
		(3.1608, 0)
		(3.4594, 0)
		(3.7644, 0)
		(4.0746, 0)
		(4.3891, 0)
		(4.7070, 0)
		(5.0278, 0)
		(5.3509, 0)
		(5.6758, 0)
		(6.0022, 0)
		(6.3297, 0)
		(6.6582, 0)
		(6.9875, 0)
		(7.3173, 0)
		(7.6476, 0.0025)
		(7.9784, 0.0221)
		(8.3094, 0.1023)
		(8.6406, 0.2718)
		(8.9721, 0.4928)
		(9.3037, 0.7007)
		(9.6354, 0.8530)
		(9.9672, 1)
	}; \addlegendentry{Graph-based Algorithm}
	
	\addplot[color=bblue, mark = star, mark options = {scale = 0.8, fill = blue}, line width = 0.8pt, style = dashed] coordinates 
	{
		(1.1756, 0)
		(1.3701, 0)
		(1.5827, 0)
		(1.8122, 0)
		(2.0574, 0)
		(2.3165, 0)
		(2.5878, 0)
		(2.8698, 0)
		(3.1608, 0)
		(3.4594, 0)
		(3.7644, 0)
		(4.0746, 0)
		(4.3891, 0)
		(4.7070, 0)
		(5.0278, 0)
		(5.3509, 0)
		(5.6758, 0)
		(6.0022, 0)
		(6.3297, 0)
		(6.6582, 0)
		(6.9875, 0)
		(7.3173, 0)
		(7.6476, 0.0025)
		(7.9784, 0.0221)
		(8.3094, 0.1023)
		(8.6406, 0.2718)
		(8.9721, 0.4928)
		(9.3037, 0.7007)
		(9.6354, 0.8530)
		(9.9672, 1)
	}; \addlegendentry{Exhaustive Search}
	
	\addplot[color=black, mark = triangle*, mark options = {scale = 1.5, fill = ggreen}, line width = 1pt] coordinates 
	{
		(1.1756, 0)
		(1.3701, 0)
		(1.5827, 0)
		(1.8122, 0)
		(2.0574, 0)
		(2.3165, 0)
		(2.5878, 0)
		(2.8698, 0)
		(3.1608, 0)
		(3.4594, 0)
		(3.7644, 0)
		(4.0746, 0)
		(4.3891, 0)
		(4.7070, 0)
		(5.0278, 0.0001)
		(5.3509, 0.0005)
		(5.6758, 0.0019)
		(6.0022, 0.0031)
		(6.3297, 0.0064)
		(6.6582, 0.0115)
		(6.9875, 0.0187)
		(7.3173, 0.0322)
		(7.6476, 0.0597)
		(7.9784, 0.1143)
		(8.3094, 0.2230)
		(8.6406, 0.4017)
		(8.9721, 0.6218)
		(9.3037, 0.7983)
		(9.6354, 0.9032)
		(9.9672, 1)
	}; \addlegendentry{Greedy Algoritm}
	
	\addplot[color=black, mark = square*, mark options = {fill = yyellow, solid}, line width = 1pt] coordinates 
	{
		(1.1756, 0.0010)
		(1.3701, 0.0021)
		(1.5827, 0.0041)
		(1.8122, 0.0072)
		(2.0574, 0.0134)
		(2.3165, 0.0235)
		(2.5878, 0.0386)
		(2.8698, 0.0619)
		(3.1608, 0.0946)
		(3.4594, 0.1395)
		(3.7644, 0.2033)
		(4.0746, 0.2771)
		(4.3891, 0.3669)
		(4.7070, 0.4581)
		(5.0278, 0.5505)
		(5.3509, 0.6461)
		(5.6758, 0.7312)
		(6.0022, 0.8028)
		(6.3297, 0.8604)
		(6.6582, 0.9041)
		(6.9875, 0.9383)
		(7.3173, 0.9641)
		(7.6476, 0.9815)
		(7.9784, 0.9897)
		(8.3094, 0.9936)
		(8.6406, 0.9970)
		(8.9721, 0.9991)
		(9.3037, 0.9996)
		(9.6354, 0.9999)
		(9.9672, 1)
	}; \addlegendentry{Random Algorithm}
	
	\addplot[color=brown, mark options = {fill = yyellow}, line width = 1pt, style = densely dashed] coordinates 
	{
		(1.1756, 0)
		(1.3701, 0)
		(1.5827, 0)
		(1.8122, 0)
		(2.0574, 0)
		(2.3165, 0)
		(2.5878, 0)
		(2.8698, 0)
		(3.1608, 0)
		(3.4594, 0)
		(3.7644, 0)
		(4.0746, 0)
		(4.3891, 0)
		(4.7070, 0)
		(5.0278, 0)
		(5.3509, 0)
		(5.6758, 0)
		(6.0022, 0)
		(6.3297, 0)
		(6.6582, 0)
		(6.9875, 0)
		(7.3173, 0)
		(7.6476, 0)
		(7.9784, 0.0001)
		(8.3094, 0.0146)
		(8.6406, 0.1208)
		(8.9721, 0.3712)
		(9.3037, 0.6350)
		(9.6354, 0.8252)
		(9.9672, 1)
	}; \addlegendentry{Exhaustive Search \\ w/o constraints}
	\end{axis}
	\end{tikzpicture}
	\caption{Cumulative Distribution Function}
	\vspace{-0.1cm}
\end{figure}

\section{Conclusion}
We have presented a novel resource allocation algorithm for V2V communications considering conflict constraints. We were able to transform the original problem into a simplified form. In our future work, we will consider ($i$) power control and ($ii$) the assumption that a subset of vehicles may belong to more than one cluster simultaneously.

\bibliography{references1} 
\bibliographystyle{unsrt}

%
%
%
%
%

\end{document}